\documentclass[twocolumn,superscriptaddress,pre]{revtex4}
\usepackage{epsfig}
\usepackage{amssymb,amsfonts,amsmath}
\usepackage[pdftex]{pstricks}
\usepackage{bm}

\include{epsf}

\newcommand{\beq}{\begin{equation}}
\newcommand{\eeq}{\end{equation}}
\newcommand{\bea}{\begin{eqnarray}}
\newcommand{\eea}{\end{eqnarray}}

\usepackage{amsmath}
\usepackage{amssymb}
\usepackage{amsthm}
\usepackage{amsfonts}
\usepackage{soul}
\usepackage[normalem]{ulem}

\let\a=\alpha   \let\g=\gamma  \let\d=\delta 
     \let\l=\lambda

\let\la = \langle \let\ra = \rangle

\begin{document}

\title{Separating intrinsic interactions from extrinsic correlations\\
 in a network of sensory neurons}

\author{Ulisse Ferrari}
\thanks{Correspondence should be sent to \url{ulisse.ferrari@gmail.com}.}
\affiliation{Sorbonne Universit\'e, INSERM, CNRS, Institut de la Vision, 17 rue Moreau, 75012 Paris, France.}
\author{St\'{e}phane Deny}
\affiliation{Neural Dynamics and Computation Lab, Stanford University, California}
\author{Matthew Chalk}
\affiliation{Sorbonne Universit\'e, INSERM, CNRS, Institut de la Vision, 17 rue Moreau, 75012 Paris, France.}
\author{Ga\v{s}per Tka\v{c}ik}
\affiliation{Institute of Science and Technology,  Klosterneuburg, Austria}
\author{Olivier Marre}
\affiliation{Sorbonne Universit\'e, INSERM, CNRS, Institut de la Vision, 17 rue Moreau, 75012 Paris, France.}
\affiliation{Equal contribution}
\author{Thierry Mora}
\affiliation{Laboratoire de physique statistique, CNRS, Sorbonne Universit\'e, Universit\'e Paris-Diderot and \'Ecole normale sup\'erieure (PSL), 24, rue Lhomond, 75005 Paris, France}
\affiliation{Equal contribution}

\begin{abstract}

Correlations in sensory neural networks have both extrinsic and intrinsic origins. Extrinsic or stimulus correlations arise from shared inputs to the network, and thus depend strongly on the stimulus ensemble. Intrinsic or noise correlations reflect biophysical mechanisms of interactions between neurons, which are expected to be robust to changes of the stimulus ensemble. Despite the importance of this distinction for understanding how sensory networks encode information collectively, no method exists to reliably separate intrinsic interactions from extrinsic correlations in neural activity data, limiting our ability to build predictive models of the network response. 
In this paper we introduce a general strategy to infer {population models of interacting neurons that collectively encode stimulus information}. The key to disentangling intrinsic from extrinsic correlations is to infer the {couplings between neurons}  separately from the encoding model, and to combine the two using corrections calculated in a mean-field approximation. We demonstrate the effectiveness of this approach on retinal recordings. The same coupling network is inferred from responses to radically different stimulus ensembles, showing that these couplings indeed reflect stimulus-independent interactions between neurons. The inferred model predicts accurately the collective response of retinal ganglion cell populations as a function of the stimulus.

\end{abstract}

\maketitle

\section{Introduction}

An important goal in sensory neuroscience is to build network models to understand how sensory stimuli are encoded by the collective activity of neuronal populations. 
Pioneering work initiated in the retina \cite{Schneidman03,Schneidman03b,Schneidman06,Shlens06} proposed to use disordered Ising models to characterize the joint activity of neurons in early sensory systems \cite{Cocco09,Ganmor11,Tkacik13a,Tkacik13b,Tkacik14}, sensory cortices \cite{Marre09} and beyond \cite{Hamilton13,Tavoni17,Nghiem18,Meshulam17}.
Motivated by the principle of Maximum Entropy \cite{Jaynes82}, these models represent neurons as binary spins (spike or silence), biased by local fields and interacting through a network of pairwise couplings. The model parameters are fitted to reproduce the empirical mean neuron activities and pairwise correlations between them.

A major limitation of this approach in the context of sensory systems is that the inferred couplings are only effective and do not directly follow network circuitry. Instead, they reflect two different sources of correlations. On one hand, two neurons can be correlated because they receive correlated or common inputs. For example, in the retina, if the stimulus is correlated over space, nearby neurons will receive similar inputs and consequently will respond synchronously. This type of correlation has been termed `signal correlation' and strongly depends on the actual stimulus and its statistics. On the other hand, neurons might be correlated because of actual interactions in the neural network, either because they are connected directly through gap junctions \cite{Bloomfield09,Volgyi09} or indirect pathways \cite{Brivanlou98}, or because they receive the same noise sources from photoreceptors \cite{Ala11}. These types of correlations have been termed `noise correlations' and results from the physological wiring of the network \cite{Brivanlou98}.
Similar network effects are present in all sensory systems \cite{Smith08,Cohen11}.
When fitting Ising models, signal and noise correlations are mixed together and difficult to disentangle.
Consequently, the inferred couplings do not only reflect properties of the network circuitry, but also incidental properties of the actual stimulus.

The inability to separate extrinsic from intrinsic correlations limits the interpretability of these models and their capacity to generalize across {different} conditions. 
For example, if {two Ising} models are trained on the neural responses to two different types of stimuli, their interactions terms will be different \cite{Cocco09}. The model thus cannot generalize and fails to predict the collective behaviour in response to a different type of stimulus. Interpreting a change in the interaction terms will also be difficult: it could trivially reflect changes in the stimulus statistics, or could correspond to changes in the network of couplings, and the way the network processes stimuli (adaptation). Modeling the influence of the stimulus is therefore crucial to understand the collective processing performed by sensory networks. 
Models with stimulus-dependent fields and couplings have been proposed to describe stimulus correlations and network effects \cite{Granot-Atedgi13,Tkacik13a,Pillow08}.
However, here we argue that the strategy proposed to fit these models from neural recordings does not ensure a proper disentanglement between these two sources of correlated activity. Thus, a general strategy to model and accurately separate stimulus and noise correlations in neural networks is still lacking.

Here we propose a general method to achieve this task. 
{We define a population encoding model where each neuron's spiking probability is governed by its couplings to other neurons, and by an external, time-dependent field encoding the effect of the stimulus.}
We describe a working strategy for learning the parameters of this model from neural recordings. First, we infer the coupling matrix from population responses to repetitions of short films. Second, we model how each neuron's firing rate depends on the stimulus, with no regard for noise correlations. Third, we use a mean field (Thoughless-Anderson-Palmer) approximation to calculate the value of the fluctuating field as a function the stimulus from those predicted rates, corrected by the influence of the network.

{We apply our model to describe the responses of retinal neurons to a visual stimulus.} We quantify the importance of noise correlations in this system, and we fit the corresponding coupling matrix from recordings of responses to repeated films. We then combine this coupling matrix with a previously proposed model of stimulus encoding fitted on single cells \cite{Deny17}
to obtain a complete model that reproduces the population response. We show that this strategy can be used to obtain accurate predictions across different stimulus statistics. 
We have therefore found a way to design and train models of population responses that can generalize across different stimulus ensembles. The method can be applied to any system where a single-cell encoding model is available.

\section{A general population model}
\subsection{Model definition}
We start by introducing a general model of the activity of a population of correlated neurons, labeled by $i=1,\ldots,N$, in response to a stimulus. 
This probabilistic model accounts for both arbitrary single-cell dependences on the stimulus, and direct interactions between cells. 
Let us denote by $n_i$ the number of spikes emitted by cell $i$ during a short time bin. 
The probability distribution of spiking patterns ${\bf n}=(n_1,\ldots,n_N)$ in response to a time-dependent stimulus $S$ is given by:
\begin{equation}
P_\textrm{pop}({\bf n}|t)= \frac{1}{Z}\exp\left[-H_1({\bf n},t)-H_2({\bf n})\right],\label{eq:SDBM}
\end{equation}
where $Z$ is a normalization constant  and 
\begin{eqnarray}
H_1({\bf n},t)&=&-\sum_{i=1}^N \left[h_i(t)  n_i - \g\, n_i^2 - \d\,  n_i^3 - \ln n_i!\right]\\
H_2({\bf n})&=&-\sum_{i\leq j} J_{ij} n_i n_j.
\end{eqnarray}
{$H_1$ and $H_2$ encode extrinsic and intrinsic sources of correlations in the population, respectively.}
The first term $H_1$ accounts for the behaviour of single cells in response to the stimulus.
$h_i(t)=\hat h_i[S_t]$ corresponds to a time-dependent external field applied to neuron $i$, which reflects the influence of the past stimulus $S_t$ at time $t$. The functional form and parametrization of $\hat h_i$ as a function of the stimulus will be prescribed later, and depend on the particular sensory system and stimulus of interest. The quadratic, cubic, and factorial terms in $n_i(t)$ in $H_1$ correspond to a correction to the Poisson distribution of spikes allowing for general dependencies between the mean and variance of $n_i$. These corrections have been shown to be essential for describing single neurons \cite{Ferrari18a}. 
The second interaction term $H_2$  is parametrized by a matrix of couplings between neurons,
${\bf J}=(J_{ij})$. 

Given recordings of the activity of a neural population presented with a known sensory stimulation, the goal is to infer the parameters of the model to best predict the collective response to arbitrary stimuli.

\subsection{Why direct likelihood maximization cannot be used}
\label{inference}
Our goal is to infer couplings that solely reflect noise correlations, for two reasons. First, this makes the values of the different parameters easier to interpret. Second, while the stimulus correlation will systematically change with the stimulus statistics, noise correlations may reflect some intrinsic network properties, and the corresponding couplings should be robust to the stimulus statistics. By separating the two types of correlations, we want to develop models that can generalize and predict responses to stimulus ensembles that are radically different from the ones they were trained on. 

One strategy to infer the population model \eqref{eq:SDBM} could be to estimate all the parameters by maximizing the likelihood on the complete dataset. However, this approach does not explicitly separate noise and signal correlations in the data, and is expected to misestimate the coupling matrix in the inference procedure. This effect comes from the fact that the stimulus-encoding model is never perfect. As a result, when the stimulus-dependent fields $h_i(t)$ fail to perfectly reproduce single neuron activities, the interaction field $\sum_j J_{ij}\sigma_j$ may try to compensate this error by drawing additional information about the stimulus from the activity of other neurons, instead of reserving these couplings for accounting for noise correlations. Previous work \cite{Granot-Atedgi13} combining a simple (i.e. linear) stimulus-encoding model with neuron-neuron couplings, and using maximum likelihood as an inference method, shows a clear example of this effect: the addition of couplings improves the prediction of single-cell firing rates (Fig. 2b of \cite{Granot-Atedgi13}).
An extreme version of this phenomenon is also demonstrated in \citep{Tkacik14}, Fig.~15, where the response of one neuron to a natural scene is predicted from the responses of the other neurons without any use of the stimulus. 

It should also be noted that the full maximum-likelihood task can be computationally hard. The inference of precise single-cell encoding models such as described in \cite{Deny17} and the inference of complete interaction networks are each computationally costly, and combining both in a single maximum-likelihood maximization would require developing new methods and algorithms.

To demonstrate the effect described above in a computationally tractable case, we inferred a Generalized Linear Model (GLM) to our retinal data (see Sec.~\ref{sec:data} for a description of the data). 
The GLM is arguably the 
{most popular} model to account for some dependence on the stimulus and for noise correlations using linear filters \cite{Pillow08}. Note that the GLM is not expected to accurately describe retinal activity because of its linear assumptions \cite{Deny17}, but it can still be used to make our point that its inference mixes signal and noise correlations. As shown in App.~\ref{App:GLM}, straightforward likelihood maximization does not guarantee that coupling terms only reflect noise correlations; rather, they reflect an uncontrolled combination or mixture of signal and noise correlations, which have very different biological interpretations.

In summary, from previous observations and the analysis on our own data with the GLM, we conclude that direct maximum-likelihood optimization of the model's parameters is not appropriate and will give biased estimates of the model parameters, and in particular of the coupling network.

\subsection{Inference of neuronal couplings}\label{sec:couplings}
In order to obtain the parameters values $J_{ij}$ that account solely for noise correlations, we need a model that reproduces perfectly the time course of the firing rate of each neuron, thereby setting aside the question of its dependence on the stimulus.
To construct such a model, we use repetitions of the same stimulation to estimate the empirical firing rate of cell $\lambda_i(t)$ (the average number of spikes in each time bin across repetitions). We then let the time-dependent fields $h_i(t)$ in the population model {(\ref{eq:SDBM})} be inferred from the responses to the repeated stimulus along with the couplings $J_{ij}$, by maximizing the likelihood following \cite{Granot-Atedgi13}. In the maximum-likelihood fit, the fields $h_i(t)$ act as Lagrange multipliers (or chemical potentials if the spike count is viewed as a particle number) enforcing the value of firing rate of each neuron in each time bin, $\la n_i(t)\ra = \lambda_i(t)$, while the couplings $J_{ij}$ enforce the noise correlations averaged over the repeated stimulus: $(1/T)\sum_{t=1}^T \la (n_i(t)-\lambda_i(t)) (n_j(t)-\lambda_j(t))\ra$. 
{Thanks to these constraints, the time courses of the firing rates are exactly reproduced by construction, and so are the resulting stimulus correlations.
The inference of the couplings thus only reflects noise correlations.}

Note that the model inferred in this way is not a 
{stimulus} encoding model. It cannot predict the spiking activity in response to a different stimulus sequence than the one used in the repetition. To do this, we need to learn how $h_i(t)$ depends on the stimulus, which is the object of the next step of the inference.

\subsection{Conditionally independent model}\label{sec:condind}
Let us now assume that a model can be built to predict the firing rate $\l_i(t)$ of each neuron as a function of the presented stimulus, $\l_i(t) =\hat \l_i[S_t]$.
This step depends on the specific sensory system studied, as well as on the stimulus ensemble. We will see an explicit example of such as model in the case of the retina in Sec.~\ref{sec:condindretina}.

From these firing rate predictions, we derive a non-interacting model of neurons in which the interaction term $H_2$ has been removed:
\begin{equation}\label{eq:condind}
P_\textrm{CI}({\bf n}|t)= \frac{1}{Z}\exp\left[-H_1({\bf n},t)\right],
\end{equation}
where the fields are set as a function of the stimulus, $h_i(t)=\hat h^0_i[S_t]$, to enforce the constraint $\la n_i(t)\ra_{\rm CI} = \hat \l_i[S_t]$. 
This model is conditionally independent, meaning that neurons respond independently of each other, when conditioned on a given stimulus.

\subsection{Putting it together: mean-field correction to network effects}
\label{sect:meanfield}
{Now that we have inferred the parameters of the interaction network $(J_{ij})$ and of the conditionally independent model (the functions $\hat h^0_i$), the last step is to combine them to obtain the complete population encoding model \eqref{eq:SDBM}. In doing so, we must be careful to correct for the effect of the network on the activity of each neuron. 

Because of the interaction term $H_2$, each cell receives an additional field $\sum_{j} J_{ij} n_j$ from the rest of the network. This field is stochastic and explains noise correlations.
However, it also generates a mean contribution $\Delta \hat h_i[S_t]$ that affects the cell firing rate. This contribution thus needs to be removed from the stimulus-dependent field: $\hat h_i[S_t] = \hat h^0_i[S_t]- \Delta \hat h_i[S_t]$.

To estimate this correction, we compute the Thouless, Anderson and Palmer (TAP) free energy formalism of the model (\ref{eq:SDBM}) \cite{Thouless77} and use it to derive an approximation for $\Delta \hat h_i$.
We followed \cite{Kappen97} and \cite{Tanaka98} and apply a second-order Plefka expansion \cite{Plefka82} (see App.~\ref{appMFT} for more details).
The result is:
\begin{eqnarray}
\Delta \hat h_i &\approx&  \sum_{j\neq i} J_{ij} \hat \l_j + J_{ii} \Big(  V'\big(\hat \l_i\big) + 2 \hat \l_i \Big) \nonumber \\
&& + \frac{1}{2} V'\big(\hat \l_i\big)  \sum_{j\neq i} J^2_{ij} V\big(\hat \l_j\big)  + \frac{1}{2} J^2_{ii} W'(\hat \l_i\big) \label{eq:mfH_Binomial}
\end{eqnarray}
where the first two terms are the mean-field contributions of network and self couplings, whereas the lasts two are their Onsager \cite{Onsager36} reaction terms.
In (\ref{eq:mfH_Binomial}) we have introduced the functions:
\begin{eqnarray}
V(\lambda) &\equiv& \la n^2\ra_{\lambda} - \la n\ra_{\lambda}^2\\
W(\lambda) &\equiv& \la n^4\ra_{\lambda} - \la n^2\ra_{\lambda}^2
                    - \frac{\big( \la n^3\ra_{\lambda}-\la n^2\ra_{\lambda} \la n\ra_{\lambda} \big)^2 }{V(\lambda)} \label{eq:W}
\end{eqnarray}
{where $\la \cdot\ra_\lambda$ denote averages according to the distribution:
\begin{equation}
P_\lambda(n)=\frac{1}{Z}\exp\left[h(\lambda)n -\gamma n^2-\delta n^3-\ln(n!)\right]
\end{equation}
where $h(\lambda)$ is set to enforce $\la n\ra =\lambda$.
Note that the correction (\ref{eq:mfH_Binomial}) only depends on the stimulus through the predicted firing rates ${\bf \hat \l}[S_t]$, and not on the spike counts ${n_i}(t)$, as desired.
}

This computation completes the procedure. The population model \eqref{eq:SDBM}, now endowed with stimulus-dependent field functions $\hat h_i[S_t]$, spike count parameters $\gamma$ and $\delta$, and network couplings $J_{ij}$, can be used to predict the population response to any stimulus $S$. Next we apply the procedure to the retinal network.

\begin{figure*}[ht]
\begin{center}
\includegraphics[clip=true,keepaspectratio,angle=-0,width=2.0\columnwidth]{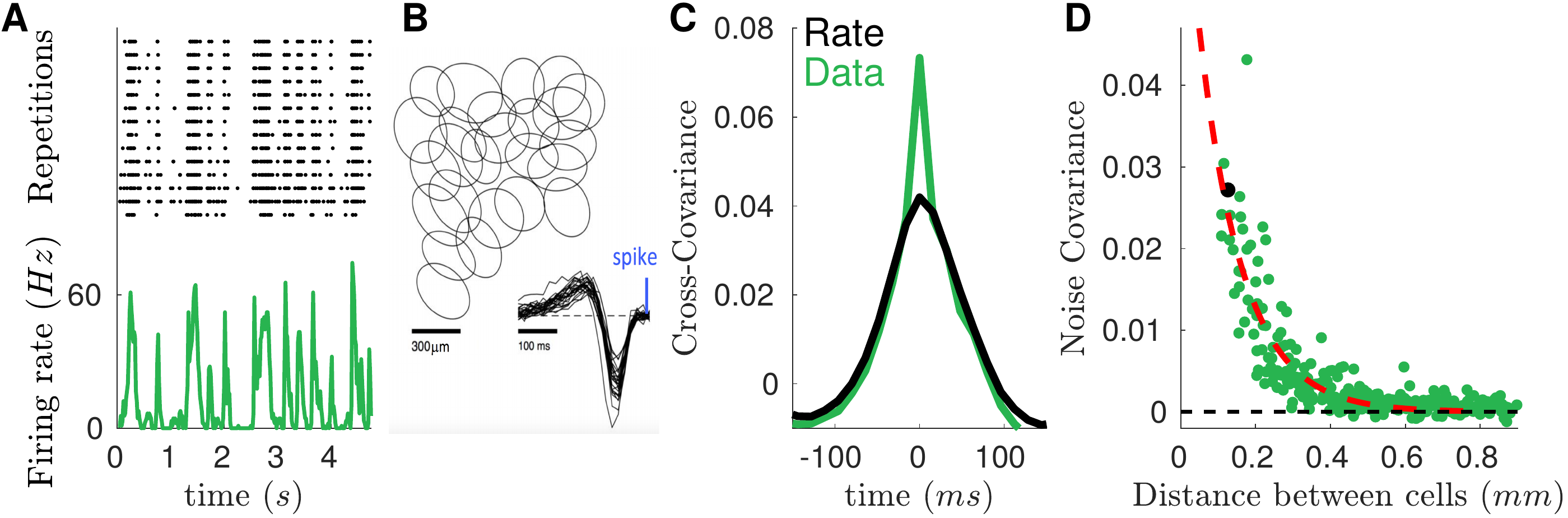}
\end{center}
\caption{\textbf{\textit{Ex-vivo} recording of retinal ganglion cell shows short time-scale noise correlations.}
A) Top: activity of one ganglion cell over repeated presentations of the same stimulus. Bottom: firing rate of the cell across time.
B) Receptive field mosaic of the isolated OFF ganglion cells
C) Empirical cross-covariance between two example cells (green trace) superimposed on the covariance between their average firing rate (black trace). 
{The difference, which is large only at short time-scales, is the noise correlation.}
D) Zero-lag noise correlation as a function of cell pair distance. 
Black point refers to the pair of neurons plotted in panel C.
Red dashed line shows an exponential fit with spatial scale of $0.11mm$.
}
\label{fig:rawData}
\end{figure*}

\section{Application to the retina}
\subsection{Description of the data}\label{sec:data}
We reanalyzed a dataset of \textit{ex-vivo} recording of the activity of retina ganglion cells (RGCs) in Long-Evans adult rats using a multi-electrode array \cite{Deny17} (see Fig.~\ref{fig:rawData}A, top for an example response). 
The stimulus was a movie of two parallel bars, whose trajectories followed the statistics of two independent overdamped stochastic oscillators.
Additionally, a white noise stimulus (random binary checkerboard) was projected for one hour to allow for receptive field estimation and RGC type identification. Raw voltage traces were stored and spike-sorted off-line through a custom spike sorting algorithm \cite{Yger16}. We applied a standard clustering analysis based on the cell response to various stimuli to isolate a population of OFF ganglion cells of the same type. Their receptive fields tiled the visual field to form a mosaic (Fig.~\ref{fig:rawData}B). 

The ganglion cell spike times were binned in $(1/60)$s time windows (locked to the stimulus frame rate) to estimate the empirical spike counts, $n_i(t)$ for cell $i$ in time bin $t$. 
The stimulus alternates between non-repeated sequences of random bar trajectories, and a repeated sequence of randomly moving bars, displayed 54 times. 
Non-repeated sequences were divided into training (two thirds of the total) and testing (one third) sets.
Repeated sequences were equally divided into training and testing sets by splitting the repeated trajectory in two different halves.

\subsection{Stimulus and noise correlations}
Before describing the inference of the model, we first briefly characterize the amount and properties of the stimulus and noise correlations.
To this end, we estimate the mean firing rate $\l_i(t)$ as a function of time in response to the repeated stimulus sequence as the empirical mean of $n_i(t)$ across repetitions (Fig.~\ref{fig:rawData}A, bottom), and its temporal average as $\l_i \equiv (1/T) \sum^T_{t=1} \l_i(t)$, where $t=1,\ldots,T$ spans the duration of the repeated stimulus.
We measure the covariance between pairs of cells (example in Fig.~\ref{fig:rawData}C) computed from the repeated dataset between two cells. 
The total pairwise covariance, represented in green in Fig.~\ref{fig:rawData}C, is the sum of stimulus and noise covariances,
\begin{equation}
c^{\rm tot}_{ij}(\tau)=c^{S}_{ij}(\tau)+c^{N}_{ij}(\tau),
\end{equation}
with
\begin{align}
c^{\rm tot}_{ij}(\tau)&=\frac{1}{T}\sum_{t=1}^T \la (n_i(t)-\lambda_i)(n_j(t+\tau)-\lambda_j)\ra_{\rm rep},\\
c^{S}_{ij}(\tau) &= \frac{1}{T} \sum_{t=1}^T (\l_i(t) -\l_i)(\l_j(t+\tau) -\l_j),\\ \label{eq:stim-cov}
c^{N}_{ij}(\tau) &= \frac{1}{T} \sum_{t=1}^T \big\langle (n_i(t) -\l_i(t))(n_j(t+\tau) -\l_j(t+\tau)) \big\rangle_\text{rep},
\end{align}
where $\la \cdot\ra_{\rm rep}$ represents averages over stimulus repetitions.
The stimulus covariance, shown in black in Fig.~\ref{fig:rawData}C, can be calculated from the empirical firing rates $\l_i(t)$. The noise covariance corresponds to the difference between the total and stimulus covariance. Fig.~\ref{fig:rawData}C shows that this difference is
significantly different from zero only at zero lag ($\tau=0$), meaning noise correlations happen at a short time scale. This suggests they may be due to gap-junctions \cite{Brivanlou98,Trenholm14}. Only cells that are physically close (as measured by the distance between the neurons' receptive fields) have large noise correlations (Fig.~\ref{fig:rawData}D), and their values strongly decreased with distance.

\begin{figure*}[ht]
\begin{center}
\includegraphics[clip=true,keepaspectratio,angle=-0,width=2.0\columnwidth]{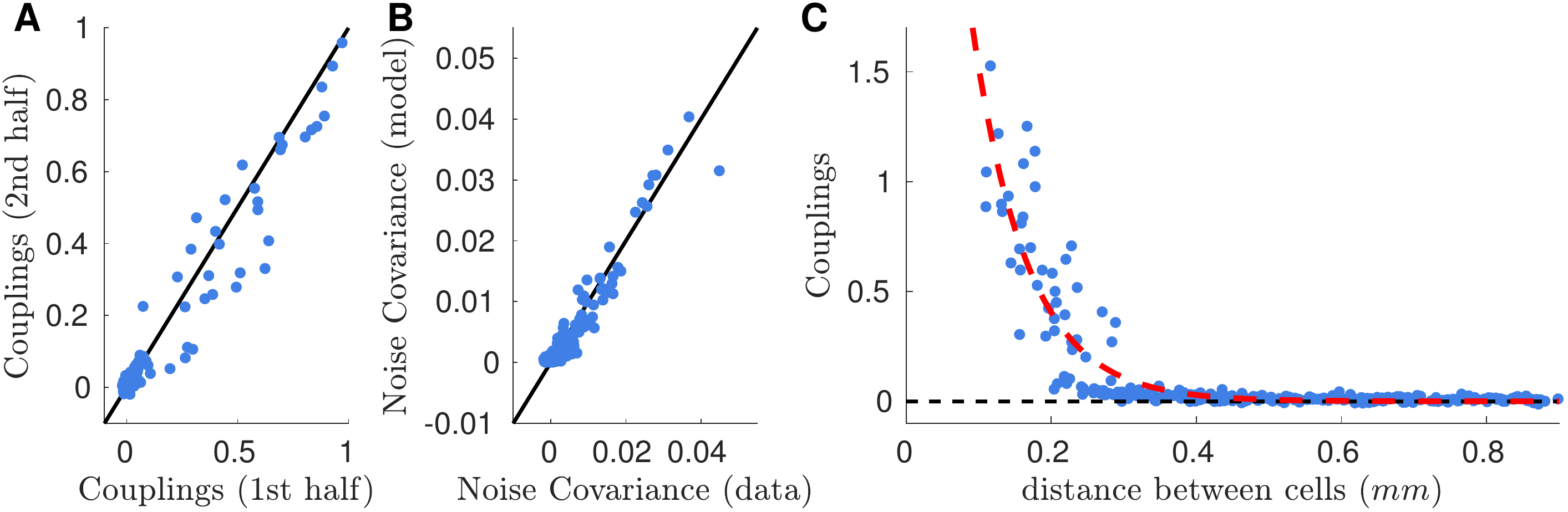}
\end{center}
\caption{\textbf{Time-dependent inference provides good estimates of the interaction network.}
A) Comparison between the inferred interaction from first (later used as training) and second (later used as testing) halves of the repeated dataset.
Black line is the identity.
B) Comparison of the predicted noise correlations when the interaction matrix ${\bm J}$  learned using the training set is applied to the testing set.
Black line is the identity.
C) The behavior with distance of the inferred interactions scales similarly to that of noise correlations (see Fig.~\ref{fig:rawData}D), although it goes to zero at shorter distances.
Red dashed line is an exponential fit with spatial constant $0.08mm$.
}
\label{fig:TDinference}
\end{figure*}

\subsection{Coupling network}\label{sec:retinacouplings}
We applied the procedure described in Sec.~\ref{sec:couplings} to the responses to the repeated stimulus to learn the coupling matrix $\bf J$.
The inference was performed by maximizing the log-likelihood (see for example \cite{Ferrari16}).
We added a small L2 penalty (with coefficient $\eta_\text{~L2} \sim 2 \cdot 10^{-6}$) on the fields $h_i(t)$ to avoid divergences to $- \infty$ when $\l_i(t)=0$.
In order to avoid spurious non-zero values, we also add an L1 penalty (with coefficient $\eta_\text{~L1} = 0.04$).
We further imposed that $J_{ii}$ be independent of $i$, for consistency with the single-cell model (see below).
}

Fig.~\ref{fig:TDinference} shows the results of the inference.
To evaluate the robustness of the inference with respect to a change of
{stimulus realization}, in Fig.~\ref{fig:TDinference}A we plot the interactions inferred from the training set against those inferred from another training set of the same size, where the bars followed a different trajectory.
The comparison shows that inferred networks are robust against a change of stimulus.

To check the validity of this approach, in Fig.~\ref{fig:TDinference}B we compare empirical noise correlations obtained with the test dataset with those predicted by the model.
To obtain this prediction, we freeze the ${\bf J}$ matrix obtained from the inference on the training set and we re-infer the $h_i(t)$ to match the firing rates of the testing set. 
The inferred coupling matrix is able to well predict the noise correlation on a part of the recording that had not been used to learn them.
Fig.~\ref{fig:TDinference}C shows the behavior of the interaction parameters as a function of the distance between the two neurons' receptive fields. 
$J_{ij}$ decreases with distance slightly faster than noise correlation (see Fig.~\ref{fig:rawData} for comparison).

\begin{figure*}[t!]
\begin{center}
\includegraphics[clip=true,keepaspectratio,angle=-0,width=1.9\columnwidth]{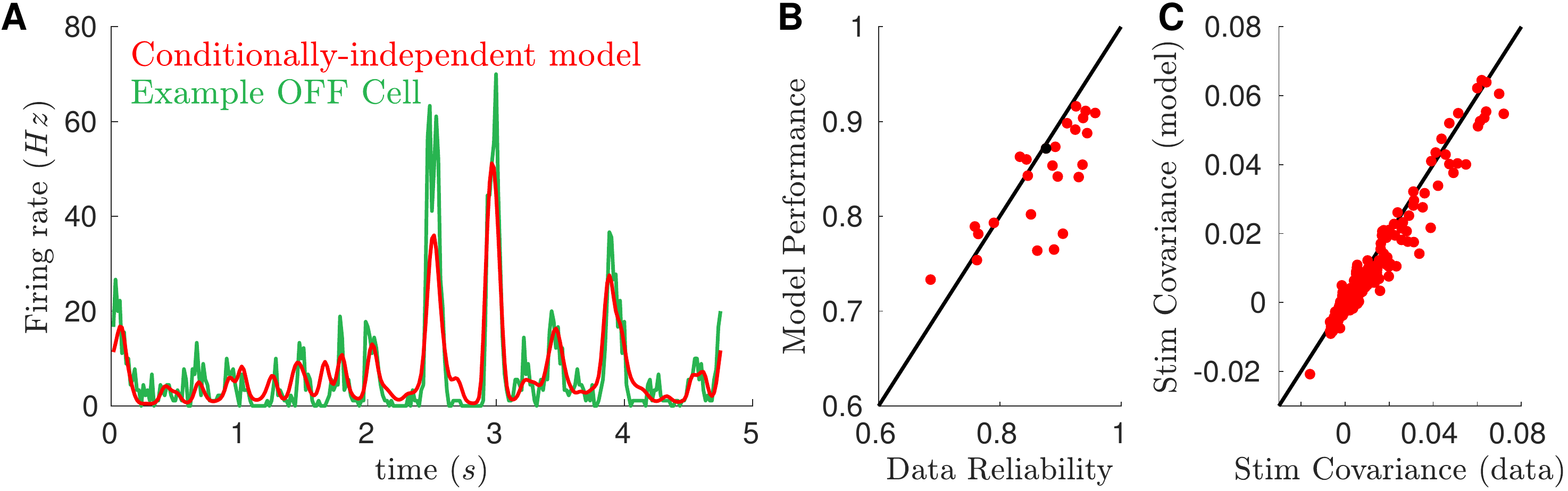}
\end{center}
\caption{\textbf{Predictions of the conditionally independent feedforward model.}
A) For an example cell, the model-predicted firing rate (red trace) is super-imposed to the empirical one (green trace). 
By computing the correlation between the two traces ($\rho=0.87$) we can estimate the model performance.
B) Scatter-plot of the model performance against the cell reliability, computed as the correlation between two halves of the repeated dataset.
Black point refers to the cell of panel A.
C) Model-predicted  stimulus covariances account for more than $84 \%$ of stimulus covariance (slope of a linear fit), but systematically under-estimate  their empirical value.
Black line is the identity.
}
\label{fig:LNLNmodel}
\end{figure*}

\subsection{A feed-forward single-cell model}\label{sec:condindretina}
To apply the procedure of Sec.~\ref{sec:condind}, we need a model for the encoding of the stimulus by single neurons. We use a previously proposed feed-forward model \cite{Deny17} that was specifically developed to predict responses to two-bar stimuli.
The stimulus $S(x,t)$, representing the time behavior of each pixel, is first convolved with a Gaussian and  biphasic factorized  kernel $K_\text{BP}(x,t)$ and then passed through rectified quadratic units with the two possible polarities:
\begin{equation}
\Theta^{\pm}(x,t)={\left[\int dx'\, dt'\,K_{\rm BP}(x-x',t-t') S(x',t')\right]}_\pm^2
\end{equation}
where $[y]_+=\max(y,0)$ {and $[y]_-=\min(y,0)$ }.
The intermediate variable $\Theta(x,t)$ is then fed into a second non-linear stage: for each cell $i$, it is first convolved with a receptive field $K^\pm_i(x,t)$ and then passed through a non-linear function:
\begin{equation}
\begin{split}
\hat \l_i[S_t] =& f_i\left(\int dx\, dt'\, K^+_i(x,t-t') \Theta^+(x,t)\right.\\
&\left.+K^-_i(x,t-t') \Theta^-(x,t) \right),
\end{split}
\end{equation}
with $f_i (y) = a_i \ln [1+ \exp \big( b_i ( y + c_i))]$.

\begin{figure*}[t!]
\begin{center}
\includegraphics[clip=true,keepaspectratio,angle=-0,width=2.0\columnwidth]{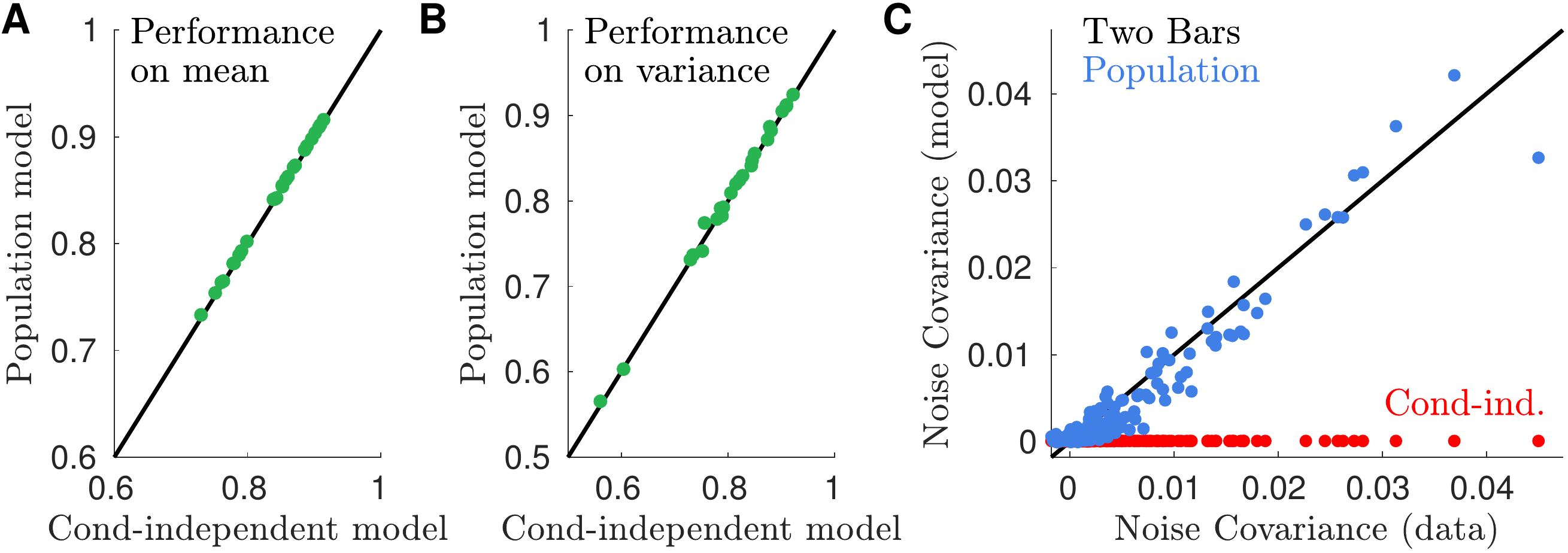}
\end{center}
\caption{\textbf{The population model predicts empirical noise covariances and performs as well as the conditional-independent model on stimulus-dependent quantities.}
A) Performance in estimating the cell firing rate (same as Fig.~\ref{fig:LNLNmodel}B) of the population model is equivalent to that of the conditionally independent model.
B) Both models also have the same performance in predicting the spike count variance.
C) The population model predicts  empirical noise covariances when applied on a testing set (blue points), while the conditionally independent model predicts zero noise covariances (red).
}
\label{fig:SDB}
\end{figure*}

To infer the parameters of the model we follow a simplified version of \cite{Mcfarland13}, where only the second non-linear stage is learned.
We keep fixed the first stage on a setting that has been shown to work efficiently \cite{Deny17}.
For the second stage we apply an iterative procedure where we maximize the log-likelihood of the data under the model given by Eq.~\ref{eq:condind}, penalized by a weighted sum of the L1 and L2 norms of the parameter vector. We used the non-repeated training set to compute the log-likelihood gradient. To avoid overfitting we early-stopped the iterative procedure when the log-likelihood computed on the non-repeated testing set stopped to increase. L1 and L2 penalties were optimized by maximizing the performance on the repeated dataset that later we will use for training the population model.

In Fig.~\ref{fig:LNLNmodel}A we compare the time course of the empirical firing rate, $\l_i(t)$, with the prediction of the inferred model for an example cell.
By computing the Pearson correlation among these two temporal traces ($\rho = 0.87$), we can estimate the performance of the model for each cell.
In Fig.~\ref{fig:LNLNmodel}B we compare this performance with the reliability of the retinal spike activity, estimated as the correlation between two disjoint subsets of responses to the repeated stimulus and found that they were comparable.
In Fig.~\ref{fig:LNLNmodel}C we show how the model predicts the empirical stimulus covariances.
Even if a small under-estimation is present, the model accounts for more than $84 \%$ of the empirical value (slope of a linear fit).

\subsection{Complete population model}\label{sec:completeretina}
The final step is to combine the single cell model and the interaction network, as explained in Sec.~\ref{sect:meanfield}.
{In Fig.~\ref{fig:SDB} we compare the performance of the population \eqref{eq:SDBM} and conditionally-independent \eqref{eq:condind} models on a testing set that was not used for learning.}
In panels A and B, we check that inferring the population model still preserves the quality of the prediction of single cell activity obtained with the independent model.
We compare the performance of the two models in reproducing the firing rate of the recorded cells (same quantity as Fig.~\ref{fig:LNLNmodel}B).
Firing rate is a stimulus dependent quantity and accordingly the two models have similar performance.
The fact that the population model's performance is not degraded compared to the single-cell model validates the approximations made to calculate the corrections ${\Delta}{\hat h}[S_t]$ within the TAP approach. 
In addition, the fact that this performance is not improved is also a positive sign: it implies that the couplings do not try to compensate for failings of the encoding model, and only reflects noise correlations.
Panel B shows that the two models have similar performance also for spike count variance.

However, the population model largely outperforms the conditionally independent model in predicting the population joint activity.
Fig.~\ref{fig:SDB}C shows how the population model accounts well for noise covariances on a testing set (blue points). 
By construction, the conditionally independent model predicts vanishing noise covariance (red points).

\subsection{Robustness of the inference of the couplings to the choice of stimulus}
\begin{figure}[th!]
\begin{center}
\includegraphics[clip=true,keepaspectratio,angle=-0,width=1.0\columnwidth]{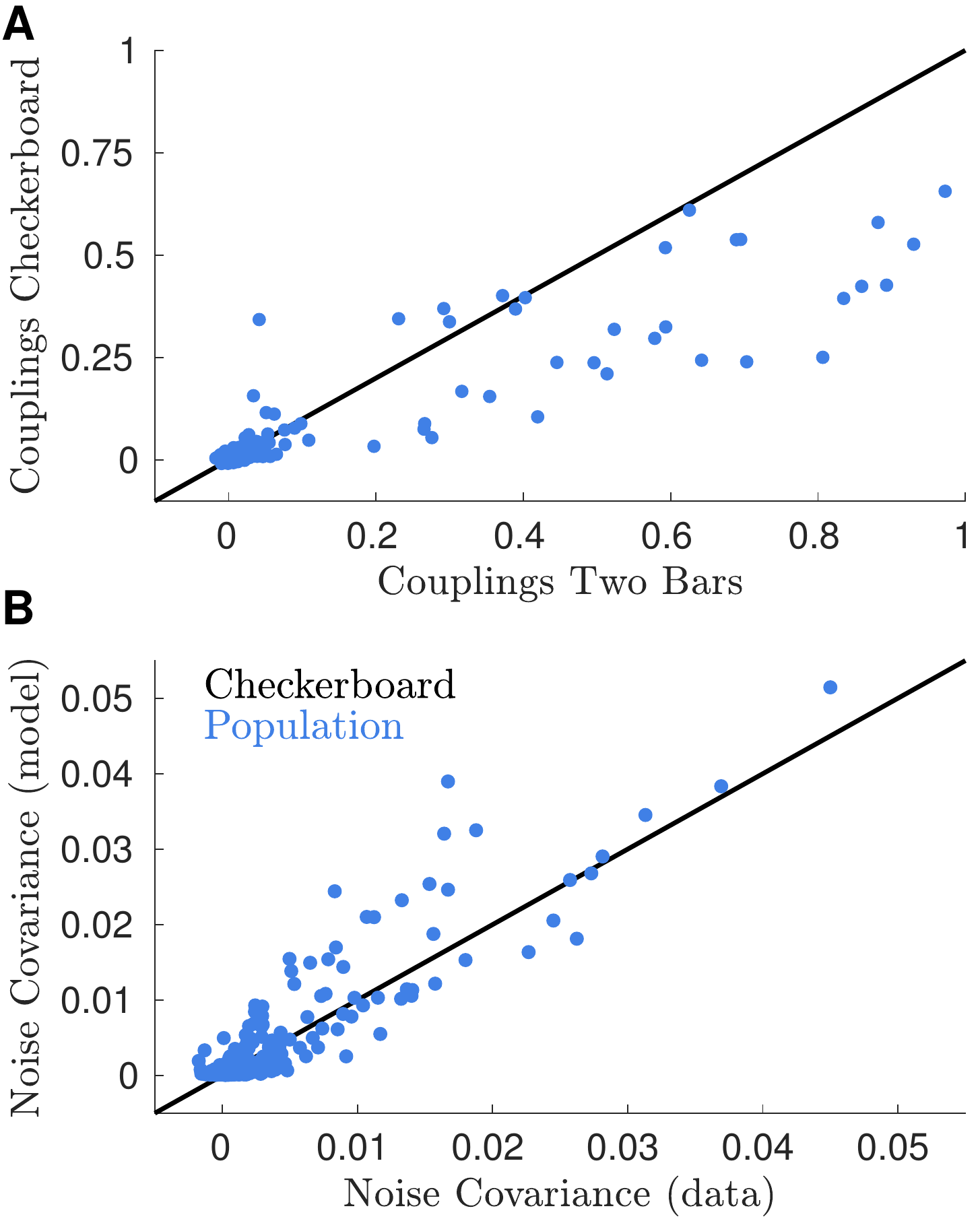}
\end{center}
\caption{\textbf{Inferred couplings generalize from one stimulus ensemble to another.}
A) Scatterplot of the inferred couplings from the response to the two bars stimulation (x-axis) against those from the response to checkerboard.
B) Population model prediction of empirical noise covariances in response to the two bars when the couplings are inferred from repeated checkerboard stimulation.
}
\label{fig:CHECK}
\end{figure}

A major challenge with fitting complex encoding models to neural responses is that they rarely generalize well. 
Here we ask if the interaction network can be inferred from a qualitatively different stimulus and then applied to our two-bars stimulation.
To test for this, we infer the couplings as in Sec.~\ref{sec:retinacouplings}, but on the response to repeated whitenoise (random checkerboard) stimuli.
In fig.~\ref{fig:CHECK}A we compare the couplings inferred from the whitenoise stimulus, to the couplings inferred from the two-bar stimulus.
We then use the coupling matrix ${\bf J}$ learned on the whitenoise stimulus in a complete population model for predicting the response to two-bar stimuli (fitted following Secs.~\ref{sec:condindretina} and \ref{sec:completeretina}).
In fig.~\ref{fig:CHECK}B we show that this model predicts noise covariances when applied to the two-bar testing set.

This demonstrates that our inference method allows us to generalize from one stimulus type to the other, and that the inferred couplings between neurons are invariant to the stimulus. 

\section{Discussion}

We have introduced a model for the spiking activity of population of sensory neurons responding to a stimulus, in which extrinsic and intrinsic correlations are clearly separated.
Our method is general and could be applied to other structures than the retina. 
It could also be extended to models where the influence of the stimulus on single cell activity is described by different non-linear models than illustrated here, e.g. a deep network with more layers \cite{Mcintosh16}.

Our inference strategy allows us to infer couplings between neurons that only reflect noise correlations between neurons, without the interference of stimulus effects due to artifacts of the learning procedure. Such effects can arise when the inference procedure tries to use the activity of other cells as a surrogate of the stimulus itself to improve the predictability of a given cell, compensating for non-linearities in the stimulus dependence that are unaccounted by the model. The inferred couplings thus show a weak dependence on the stimulus ensemble driving the neuronal response, in contrast to previous attempts where the stimulus ensemble had a major influence on the couplings \cite{Cocco09}.
Note that checkerboard and two-bars stimulations drive very different responses of the retina.
It is thus a remarkable result that noise correlations in the response to a complex film of moving objects can be predicted by couplings inferred from responses to white-noise stimuli.
This result can thus be seen as a first step toward the construction of a full model that accounts for large and heterogeneous stimulus ensembles.

Having models that account for noise correlations is a crucial first step to study their impact on coding complex stimuli. 
This impact can be quantified by comparing the population model to the conditionally independent model ($J_{ij}=0$), or to alternative population models with different coupling matrices. 
By comparing the computations performed by these models, future works will be able to understand how these fast noise correlations affect coding in the retina. 
The same strategy can be used in other sensory structures or any other noisy input-output systems. 

\section{Acknowledgements} 
We like to thank C. Gardella for useful discussion.
This work was supported by ANR TRAJECTORY, the French State program Investissements d'Avenir managed by the Agence Nationale de la Recherche [LIFESENSES: ANR-10-LABX-65], a EC grant from the Human Brain Project (FP7-604102)), NIH grant U01NS090501 and AVIESAN-UNADEV grant to OM.

\appendix
\section{Generalized linear model analysis}
\label{App:GLM}
 % SI_glm
We fitted a generalized linear model (GLM) to the responses of $N_{r}=25$
off-cells in response to the moving bar stimuli. We discretized the
response and stimulus using temporal bins of length $1.667ms$ (i.e. 600Hz;
10 times smaller than the temporal bins used for the model described
in the main paper). The response of the $i^{th}$ neuron at time $t$
is described by an integer $n_{i}(t)$ denoting the number of spikes
fired by the $i^{th}$ neuron in the $t^{th}$ time bin. The spatial
location of the presented bar stimuli were discretized into $N_{x}=100$
equi-sized bins. The stimulus at time $t$ was denoted by a binary	
vector $\boldsymbol{x}_{t}$, of which the $i^{th}$ component ($x_{i}(t)$)
was set to 1 if one of the bars was centred on the corresponding spatial location at time
$t$, and 0 otherwise. 

\begin{figure*}[t!]
\begin{center}
\includegraphics[clip=true,keepaspectratio,angle=-0,width=2.0\columnwidth]{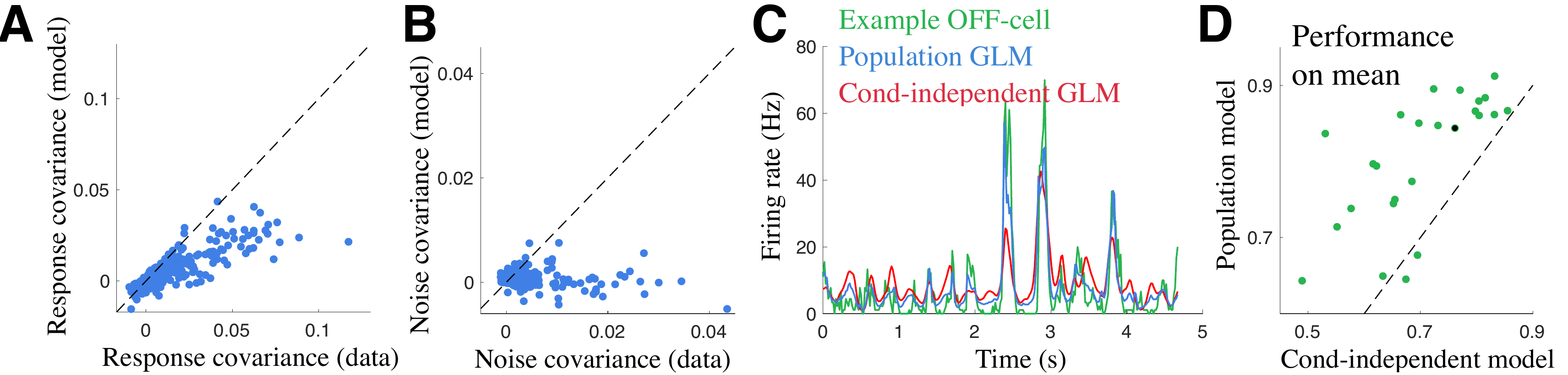}
\end{center}
\caption{(A) Total response covariances predicted by the coupled GLM model (y-axis) versus the
response covariances measured from the data (x-axis). Each point corresponds to a single neuron pair. 
(B) Noise covariances predicted by the coupled GLM model (y-axis) versus the
noise covariances measured from the data (x-axis). (C) Firing rate of a
single example cell (green; also plotted in Fig 2A), alongside the predictions of the coupled (`population') 
(blue) and uncoupled (`conditionally-independent') (red) GLM models.
(D) Correlation coefficient between the firing rate predicted by the
coupled (y-axis) and uncoupled (x-axis) GLM models and data, for each
cell. The cell shown in panel C is indicated in black. }
\label{fig:GLM_fig}
\end{figure*}

In a GLM, the spiking response of each neuron 
is assumed to be Poisson distributed, with the mean number of spikes of the $i^{th}$ neuron at 
time $t$ given by $\exp\left(r_{i}(t)\right)$, where:
\begin{equation}
r_{i}(t)=\sum_{j=1}^{N_{x}}\sum_{k=0}^{N_{w}-1}w_{ijk}x_{j}\left(t-k\right)+\sum_{j=1}^{N_{r}}\sum_{k=1}^{N_{v}}v_{ijk}n_{j}\left(t-k\right)+b_{i}.\label{eq:glm_rate}
\end{equation}
In this equation, $b_{i}$ denotes a constant bias term, $w_{ijk}$
is an element of the temporal stimulus filter (of size $[N_{r},N_{x},N_{w}]$)),
and $v_{ijk}$ is an element of the recurrent filter (of size $[N_{r},N_{r},N_{v}]$)). We used a stimulus
filter of length $N_{w}=200$ (i.e.\ $\sim$300ms), and recurrent filters
of length $N_{v}=15$ (i.e.\ $\sim$25ms).

Model parameters were fitted by maximizing an objective function consisting
of the log-likelihood (which, for a Poisson distribution, is given
up to constant by $L=\sum_{i,t}n_{i}(t)r_{i}(t)-e^{r_{i}(t)}$), minus an L2 norm
regularization term that promoted smooth filters and thus prevented
overfitting. The regularization parameters were chosen to maximize
the log-likelihood on test data, held out during model fitting. Further,
to reduce the number of parameters, and thus further reduce over\textendash fitting
we assumed that the stimulus filter could be described as the sum
of $N_{rank}=3$ spatio-temporally separable filters, each given by
a spatial filter multiplied by a temporal filter (i.e. $w_{ijk}=\sum_{l=1}^{N_{rank}}u_{ij}^{l}a_{ik}^{l}$,
where $u$ and $a$ here denote the spatial and temporal filters,
respectively). Relaxing this assumption (by increasing $N_{rank}$)
did not improve the quality of the model fit.

We evaluated  the performance of the GLM model in fitting the covariances between the responses of pairs of neurons (Fig.~\ref{fig:GLM_fig}A). Interestingly, despite giving a reasonable fit of the total response covariances (Pearson correlation $= 0.87$), the model gave a poor fit of the noise covariances (compare Fig.~\ref{fig:GLM_fig}B with Fig 4C). We wondered whether this could be because the recurrent filters, $v$, did not just capture interactions between neurons, but also compensated for the inability of the feed-forward filters, $w$, to fully capture the effects of the stimulus on neural firing rates. To see if this was the case, we compared the coupled `population model' described above, and with firing rates given by equation \ref{eq:glm_rate}, with an uncoupled `conditionally-independent' model, where the recurrent filters, $v$, were set to zero. We found that the coupled GLM model resulted in improved predictions of the recorded PSTH for nearly all recorded OFF-cells, compared to the uncoupled model (Fig.~\ref{fig:GLM_fig}C-D). This suggests that, rather than just fitting the interactions between different neurons, the coupled GLM model used the recurrent filters, $v$, to improve the prediction of how each neuron responded to the stimulus. However, it remains to be seen whether further differences between the GLM and the Ising model (e.g. time-dependent recurrent filters, Poisson distributed firing rates) could also contribute to their different performances in predicting the noise covariances.

\section{Construction of the mean-field theory and Thouless-Anderson-Palmer correction}
\label{appMFT}
We are interested in computing the TAP correction to the fields $h_i(t)$ due to the addition of the coupling term ${\bf J}$, see Eq.(\ref{eq:SDBM}).
Because we are not interested in the TAP expression for couplings nor in that for covariances, we can construct the mean-field theory for a single time-bin.
Otherwise, because the couplings are constant in time we should have considered the whole model.
To apply the Plefka expansion we introduce:
\begin{equation}
F[\alpha, {\bm h , J}] \equiv - \ln \sum _{\bf n} \exp \left\{ -H_1({\bf n}) - \alpha H_2({\bf n})  \right\} \label{eq:freeEn}
\end{equation}
where we neglect the $t$ dependence of $H_1$ because here we focus on a single time-bin.
The Legendre transform of (\ref{eq:freeEn}) reads:
\begin{equation}
G[\a, {\bm \l , J}] =  \left(  \sum_i h_i \l_i + F[\a, {\bm h , J}] ~  \right|_{ {\bm h} = \tilde{ \bm h } } \label{eq:G}
\end{equation}
where $\tilde{\bm h} = \tilde{\bm h} [\a,{\bm \l}]$ is defined implicitly from:
\begin{equation}
 \left. \frac{\partial~ \Big(  \sum_i h_i \l_i  + F[\a, {\bm h , J}] \Big)~  }{\partial h_i}   \right|_{ {\bm h}  = \tilde{ \bm h } } \!\!\!\!\!\!\!\! = \l_i - \left. \la n_i \ra^{(\a)}\right|_{ {\bm h}  = \tilde{ \bm h } }  =0 \label{eq:hTilde}
\end{equation}
where $\la \dots \ra^{(\a)}$ is the average with respect to the distribution related to the free energy (\ref{eq:freeEn}).
Our goal is to expand $G(\a, {\bm \l , j})$ in power of $\a$ up to the second order. 
At first we evaluate the derivatives:
\begin{eqnarray}
G'[\a, {\bm \l , J}] &=& \la H_2({\bf n})   \ra^{(\a)} \\  
G''[\a, {\bm \l , J}] &=& -  \la H_2({\bf n})^2   \ra_c^{(\a)} \nonumber \\  
&& - \sum_g  \frac{\partial \tilde{h}_g }{\partial \a}   \la H_2({\bf n})  ( n_g - \l_g)  \ra^{(\a)}   \\ 
\frac{\partial \tilde{h}_g }{\partial \a} &=& -  \frac{\left\la H_2({\bf n})   (n_g - \l_g) \right\ra^{(\a)}}{\la  n_g^2  \ra^{(\a)}_c  }
\end{eqnarray}
where to obtain the last equality we applied the implicit function theorem to Eq. (\ref{eq:hTilde}).
The Plefka approximation consists in estimating $G[{\bm \l, J}] = G[\a=1, {\bm \l, J}] $ from the expansion around $\a=0$ evaluated at $\a=1$:
\begin{equation}
G[{\bm \l, J}] \approx G[0, {\bm \l , J}] + G'[0, {\bm \l , J}] +  \frac{1}{2}  G''[0, {\bm \l , J}]~.\label{plefkaExp}
\end{equation}
We need $G[\a, {\bm \l , J}]$ and its derivatives at $\a = 0$.
To this aim  we note that $\tilde{h}_i[\a = 0,{\bm \l}] =  \tilde{h}_i[\l_i]$, as for $\a=0$ the system units become independent and consequently $\tilde{h}_i$ depends only on $\l_i$.
For $\a=0$, in fact, the distribution over $\{ n_i \}_{i=1}^N$ factorizes over a set of single variable distributions. this allows us to compute model expectations at $\a=0$.
For future convenience, we define the moments of such distributions:
\begin{equation}
P^{(s)} \equiv  \la n^s \ra^{(\a=0)}
\end{equation}
so that the terms in the expansion become:
\begin{eqnarray}
G[0, {\bm \l , J}] &=&  \sum_i \tilde h_i \l_i + F[\a=0, {\bm \tilde h}] \\
G'[0, {\bm \l , J}] &=&  -  \sum_{i < j} J_{ij} \l_i \l_j - \sum_{i} J_{ii} P^{(2)}_i  \\
G''[0, {\bm \l , J}] &=& -  \sum_{i<j}    J_{ij}^2 \big( P^{(2)}_i - \l_i^2 \big)\big(P^{(2)}_j - \l_j^2 \big)   \nonumber  \\
&& -  \sum_{i}    J_{ii}^2\, W_i
\end{eqnarray}
where $W$ has been defined in Eq. (\ref{eq:W}).

The mean-field equation for the fields ${\bm h}$ can be easily obtained by a reverse Legendre transform of Eq.~(\ref{plefkaExp}):
\begin{eqnarray}
h_i &=& \frac{\partial G[{\bm \l, J}] }{ \partial \l_i} \\
 &=& \tilde{h}_i[ \l_i] +  \frac{\partial G'[0, {\bm \l, J}] }{ \partial \l_i} + \frac{1}{2}  \frac{\partial G''[0, {\bm \l, J}] }{ \partial \l_i} \label{eq:mfH}
\end{eqnarray} 
which provides the expression (\ref{eq:mfH_Binomial}) for the TAP correction.

\bibliographystyle{pnas}

\begin{thebibliography}{10}

\bibitem{Schneidman03}
Schneidman, E, Bialek, W,  \& Berry, M.
\newblock (2003) {\em Journal of Neuroscience} {\bf 23}, 11539--11553.

\bibitem{Schneidman03b}
Schneidman, E, Still, S, Berry, M.~J, Bialek, W,  et~al.
\newblock (2003) {\em Physical review letters} {\bf 91}, 238701.

\bibitem{Schneidman06}
Schneidman, E, Berry, M, Segev, R,  \& Bialek, W.
\newblock (2006) {\em \em Nature \em} {\bf {\bf 440}}, 1007.

\bibitem{Shlens06}
Shlens, J, Field, G.~D, Gauthier, J.~L, Grivich, M.~I, Petrusca, D, Sher, A,
  Litke, A.~M,  \& Chichilnisky, E.
\newblock (2006) {\em Journal of Neuroscience} {\bf 26}, 8254--8266.

\bibitem{Cocco09}
Cocco, S, Leibler, S,  \& Monasson, R.
\newblock (2009) {\em \em Proc. Natl. Acad. Sci. USA \em} {\bf {\bf 106}},
  14058.

\bibitem{Ganmor11}
Ganmor, E, Segev, R,  \& Schneidman, E.
\newblock (2011) {\em \em PNAS\em} {\bf {\bf 108}}, 9679--9684.

\bibitem{Tkacik13a}
Tkacik, G, Granot-Atedgi, E, Segev, R,  \& Schneidman, E.
\newblock (2013) {\em Phys. Rev. Lett.} {\bf 110}, 058104.

\bibitem{Tkacik13b}
Tkacik, G, Marre, O, Mora, T, Amodei, D, Berry, M,  \& Bialek, W.
\newblock (2013) {\em Journal of Statistical Mechanics: Theory and Experiment}
  {\bf 2013}, P03011.

\bibitem{Tkacik14}
Tkacik, G, Marre, O, Amodei, D, Schneidman, E, Bialek, W,  \& M.J, B.
\newblock (2014) {\em \em PloS Comput. Biol.\em} {\bf {\bf 10(1)}}, e1003408.

\bibitem{Marre09}
Marre, O, El~Boustani, S, Fr\'egnac, Y,  \& Destexhe, A.
\newblock (2009) {\em Phys. Rev. Lett.} {\bf 102}, 138101.

\bibitem{Hamilton13}
Hamilton, L.~S, Sohl-Dickstein, J, Huth, A.~G, Carels, V.~M, Deisseroth, K,  \&
  Bao, S.
\newblock (2013) {\em \em Neuron \em} {\bf {\bf 80}}, 1066--76.

\bibitem{Tavoni17}
Tavoni, G, Ferrari, U, Battaglia, F, Cocco, S,  \& Monasson, R.
\newblock (2017) {\em Network Neuroscience} pp. 1--27.

\bibitem{Nghiem18}
Nghiem, T.-A, Telenzuk, B, Marre, O, Destexhe, A,  \& Ferrari, U.
\newblock (2018) {\em bioRxiv} p. 243857.

\bibitem{Meshulam17}
Meshulam, L, Gauthier, J.~L, Brody, C.~D, Tank, D.~W,  \& Bialek, W.
\newblock (2017) {\em Neuron} {\bf 96}, 1178--1191.

\bibitem{Jaynes82}
Jaynes, E.~T.
\newblock (1982) {\em \em Proc. IEEE \em} {\bf {\bf 70}}, 939.

\bibitem{Bloomfield09}
Bloomfield, S \& V\"{o}lgyi, B.
\newblock (2009) {\em Nature Reviews Neuroscience} {\bf 10}, 495--506.

\bibitem{Volgyi09}
V\"{o}lgyi, B, Chheda, S,  \& Bloomfield, S.
\newblock (2009) {\em J. Comp. Neurol.} {\bf 512}, 664--687.

\bibitem{Brivanlou98}
Brivanlou, I, Warland, D,  \& Meister, M.
\newblock (1998) {\em \em Neuron \em} {\bf {\bf 20}}, 527--539.

\bibitem{Ala11}
Ala-Laurila, P, Greschner, M, Chichilnisky, E,  \& Rieke, F.
\newblock (2011) {\em Nature neuroscience} {\bf 14}, 1309--1316.

\bibitem{Smith08}
Smith, M.~A \& Kohn, A.
\newblock (2008) {\em Journal of Neuroscience} {\bf 28}, 12591--12603.

\bibitem{Cohen11}
Cohen, M.~R \& Kohn, A.
\newblock (2011) {\em Nature neuroscience} {\bf 14}, 811.

\bibitem{Granot-Atedgi13}
Granot-Atedgi, E, Tkacik, G, Segev, R,  \& Schneidman, E.
\newblock (2013) {\em PLOS Computational Biology} {\bf 9}, 1--14.

\bibitem{Pillow08}
Pillow, J, Shlens, J, Paninski, L, Sher, A, Litke, A, Chichilnisky, E.~J,  \&
  Simoncelli, E.
\newblock (2008) {\em \em Nature \em} {\bf {\bf 454}}, 995--999.

\bibitem{Deny17}
Deny, S, Ferrari, U, Mace, E, Yger, P, Caplette, R, Picaud, S, Tka{\v{c}}ik, G,
   \& Marre, O.
\newblock (2017) {\em Nature communications} {\bf 8}, 1964.

\bibitem{Ferrari18a}
Ferrari, U, Deny, S, Marre, O,  \& Mora, T.
\newblock (2018) {\em bioRxiv} p. 243543.

\bibitem{Thouless77}
Thouless, D, Anderson, P,  \& Palmer, R.
\newblock (1977) {\em \em Phil. Mag.\em} {\bf {\bf 35}}, 593.

\bibitem{Kappen97}
Kappen, H \& Rodriguez, F.
\newblock (1997) {\em \em Neural Comput. \em} {\bf {\bf 10}}, 1137--1156.

\bibitem{Tanaka98}
Tanaka, T.
\newblock (1998) {\em \em Phys. Rev. E\em} {\bf {\bf 58}}, 2302.

\bibitem{Plefka82}
Plefka, T.
\newblock (1982) {\em Journal of Physics A: Mathematical and general} {\bf 15},
  1971.

\bibitem{Onsager36}
Onsager, L.
\newblock (1936) {\em Journal of the American Chemical Society} {\bf 58},
  1486--1493.

\bibitem{Yger16}
Yger, P, Spampinato, G. L.~B, Esposito, E, Lefebvre, B, Deny, S, Gardella, C,
  Stimberg, M, Jetter, F, Zeck, G, Picaud, S, Duebel, J,  \& Marre, O.
\newblock (2016) {\em bioRxiv}.

\bibitem{Trenholm14}
Trenholm, S, McLaughlin, A.~J, Schwab, D, Turner, M, Smith, R, Rieke, F,  \&
  Awatramani, G.
\newblock (2014) {\em Nature neuroscience} {\bf 17}, 1759--1766.

\bibitem{Ferrari16}
Ferrari, U.
\newblock (2016) {\em Phys. Rev. E} {\bf 94}, 023301.

\bibitem{Mcfarland13}
McFarland, J, Cui, Y,  \& Butts, D.
\newblock (2013) {\em PLoS Comput Biol} {\bf 9}, e1003143.

\bibitem{Mcintosh16}
McIntosh, L, Maheswaranathan, N, Nayebi, A, Ganguli, S,  \& Baccus, S.
\newblock (2016) {\em Deep learning models of the retinal response to natural
  scenes}.
\newblock pp. 1361--1369.

\end{thebibliography}

\end{document}